# Direct Visualization of Spatial inhomogeneity of Spin Stripes Order in La$_{1.72}$Sr$_{0.28}$NiO$_4$

Gaetano Campi[1]*, Nicola Poccia[2]*, Boby Joseph[3], Antonio Bianconi[1,4], Shrawan Mishra[5,6], James Lee[5], Sujoy Roy[5], Agustinus Agung Nugroho[7], Marcel Buchholz[8], Markus Braden[8], Christoph Trabant[9], Alexey Zozulya[10,11], Leonard Müller[10], Jens Viefhaus[10], Christian Schüßler-Langeheine[9], Michael Sprung[10], Alessandro Ricci[10,4]*

[1]*Institute of Crystallography, CNR, via Salaria Km 29.300, Monterotondo Roma, 00015, Italy*
[2]*Institute for Metallic Materials, Leibniz IFW Dresden, 01069 Dresden, Germany*
[3]*Elettra Sincrotrone Trieste. Strada Statale 14 - km 163.5, AREA Science Park, I-34149 Basovizza, Trieste, Italy*
[4]*Rome International Center for Materials Science Superstripes RICMASS, Via dei Sabelli 119A, 00185 Roma, Italy*
[5]*Advanced Light Source, Lawrence Berkeley National Laboratory, Berkeley, California 94720, USA*
[6]*School of Materials Science and Technology, Indian Institute of Technology, Banaras Hindu University, Varanasi-221005, India*
[7]*Faculty of Mathematics and Natural Sciences Intitut Teknologi Bandung, Jl. Ganesha 10 Bandung, 40132, Indonesia*
[8]*II. Physikalisches Institut, Universität zu Köln, Zülpicher Str. 77, 50937 Köln, Germay*
[9]*Helmholtz-Zentrum Berlin für Materialien und Energie GmbH, Institute Methods and Instrumentation in Synchrotron Radiation Research, Albert-Einstein-Str. 15, 12489 Berlin, Germany*
[10]*Deutsches Elektronen-Synchrotron DESY, Notkestraße 85, D-22607 Hamburg, Germany*
[11]*European X-Ray Free-Electron Laser Facility GmbH Holzkoppel 4, 22869 Schenefeld, Germany*

**In several strongly correlated electron systems, defects, charge and local lattice distortions are found to show complex inhomogeneous spatial distributions. There is growing evidence that such inhomogeneity plays a fundamental role in unique functionality of quantum complex materials. La$_{1.72}$Sr$_{0.28}$NiO$_4$ is a prototypical strongly correlated material showing spin striped order associated with lattice and charge modulations. In this work we present the spatial distribution of the spin organization by applying micro X-ray diffraction to La$_{1.72}$Sr$_{0.28}$NiO$_4$, mapping the spin-density-wave order below the 120K onset temperature. We find that the spin-density-wave order shows the formation of nanoscale puddles with large spatial fluctuations. The nano-puddle density changes on the microscopic scale forming a multiscale phase separation extending from nanoscale to micron scale with scale-free distribution. Indeed spin-density-wave striped puddles are disconnected by spatial regions with different stripe orientation or negligible spin-density-wave order. The present work highlights the complex nanoscale phase separation of spin stripes in nickelate perovskites and opens the question of the energetics at domain interfaces.**

KEYWORDS: nanoscale phase separation, spin stripes, nickelates, superstripes, quantum complex materials

The complex organization of different orders seems to have a fundamental role in the mechanism governing the emergence of unique functionalities in quantum materials [1]. In cuprate perovskites [1-3], charge density waves [4], accompanied by local lattice distortions [5], have been found to be spatially inhomogeneous with the formation of "striped charge puddles" anti-correlated with competing puddles of "striped dopants rich clusters" [6] which shed light on the long-standing research on complexity in doped perovskites. These findings have been possible due to the ability to focus X-ray synchrotron radiation to smaller and smaller spots, allowing, in nominal single crystals, the direct visualization of the inhomogeneity in the bulk structure from nano to micro scale. In order to determine the role that the spatial distribution of ordered phases in cuprates plays for the superconductivity, it is instructive to study a non-superconducting reference system like the layered nickelates [7]. Keeping this idea in mind, we push forward the investigation of the spatial distribution of spin-density-wave stripes ordering (SDW-stripes) in the La$_{1.72}$Sr$_{0.28}$NiO$_4$ nickelates.

It is well known that stripes appear in layered nickelates [7] in the doping interval $0.15 \leq x \leq 0.5$ [7] with spin, charge and lattice modulations. The charge and magnetic stripes can be easily investigated separately in fact the charge-order superstructure scattering in La$_{2-x}$Sr$_x$NiO$_4$ always peaks at ($2\varepsilon; 0; l$) with odd $l$, whereas the magnetic scattering exhibits peaks for ($1-\varepsilon; 0; l$) with odd and even $l$ [7,8] (The notation refers to the commonly used orthorhombic unit cell). A large number of studies on the stripe order in La$_{2-x}$Sr$_x$NiO$_4$ were performed by traditional neutron scattering [9-13] probing the spin ordering and nuclear displacements with low energy neutrons, electron diffraction [14], muon spectroscopy [15] and recently advanced novel synchrotron radiation X-ray diffraction methods have been used to detecting both magnetic and charge ordering [16-21]. Resonant elastic x-ray diffraction REXS [16-18] detects spin order directly via magnetic contrast, while electron diffraction [14], and hard XRD, or nonresonant-x-ray-magnetic-scattering (NXMS) of nickelates and related magnetic materials [19-25] probe the associated tiny lattice distortions related with polaronic magneto-elastic strain effects [26-29]. The latter approach opens new perspectives to unveil open puzzles on the complexity of the nature of stripes in La$_{2-x}$Sr$_x$NiO$_4$ nickelates and to test theories on diagonal spin stripes in strongly correlated systems with multi-orbitals and polaronic degree of freedom [26-29].

The accumulated data on the La$_{2-x}$Sr$_x$NiO$_4$ system [7-21] enable to prepare a temperature doping phase diagram of this nickelate system, which is presented in **Fig. 1A**. The red area indicates the antiferromagnetic order (AFM) dominating the lower Sr-doping $x$ given by the percentage of Sr for La substitutions in the La$_2$O$_2$ plane, which is assumed to give the number of injected doped electronic holes per Ni atom in the NiO$_2$ plane [7]. The blue and the green areas correspond to the observation of charge-density-wave order (CDW-stripes) and magnetic spin-density-wave order (SDW-stripes), respectively. The SDW-stripes modulation wave-vector direction for 0.15<x<0.5 extends in real space diagonally to the Ni-O bond direction along the *b*-direction of the orthorhombic unit cell. For samples with tetragonal crystal symmetry, as the one studied here, the stripe order itself breaks the rotational symmetry of the *ab*-plane and therefore spin stripes were assumed to show two different orientations with a 90-degree rotation around the *c*-axis with equal probability. The spin stripes lead to superstructure peaks either in the neutron diffraction pattern [9-13] and X-ray diffraction [16-21]. Those at the lowest momentum transfer occur at wave-vectors ($1-\varepsilon$, $0$, $0$) in the orthorhombic lattice for SDW-stripes order, where $\varepsilon$ is a temperature dependent incommensurability value which is well separated from different charge stripe wave-vectors in the k-space. For the present investigation we used a sample with doping level *x*=0.28 and found *ε*=0.29 at 20K to be consistent with the empirical relation for spin stripes wave-vector $\varepsilon \approx x$ close to the percentage of holes per Ni sites *x* induced by Sr doping. The red dotted-line in **Fig. 1A** represents the temperature range where the sample of this work has been studied. In order to probe the spatial evolution of the SDW-stripes order in real space we scanned a micron-size X-ray beam across the sample, probing the local SDW-stripes via the corresponding intensity of the magnetic so called SDW-stripe superlattice peak at (0.71,0) in the orthorhombic (**h,k**) plane which is well separated from the so-called CDW-stripe superlattice reflection at (0.56,0). Scanning micro X-ray Diffraction (SµXRD) has been demonstrated to be a powerful tool in unraveling material inhomogeneity in superconductors at the micro and nanoscale, and has been successfully applied to the cuprate systems doped by oxygen interstitials: HgBa$_2$CuO$_{6+y}$, known as Hg1201 [4], La$_2$CuO$_{4+y}$, known as La124, [5,6] Bi$_2$Sr$_2$CaCu$_2$O$_{8+y}$, known as Bi2212 [30], YBa$_2$Cu$_3$O$_{6+y}$ known as Y123 [31-33], to iron-based superconductors [34] and to cobaltates materials [35].

Single-crystalline La$_{1.72}$Sr$_{0.28}$NiO$_4$ was grown by floating zone technique. The seed and feed rods were prepared from polycrystalline powder obtained by solid state reaction of La$_2$O$_3$, SrCO$_3$ and NiO with an excess of NiO. The reaction was performed at 1200 °C for 20 h with intermediate grinding. The rods were densified at 1500 °C for 5 h in air. Micro X-ray diffraction measurements of SDW-stripes order in the sample were carried out at beamline P10 of PETRA III (DESY, Hamburg) using an energy of 8 KeV. The scattering signal was detected at a sample to detector distance of $5\ m$ using the large horizontal scattering set-up of beamline P10 including an evacuated flight path. A PILATUS 300K detector was used to record the X-rays scattered by the sample. By employing a focused beam with a diameter of about 1 µm and translating the sample, we mapped the spatial distribution of the (*1-ε*,0,0) peak intensity over different areas of about 40×80 µm$^2$ in steps of 1 µm in both directions resulting in 3321 diffraction images.

The SDW-stripes peak is identifiable for temperatures below T$_{SDW}$ = 120K in agreement with previous works [7]. On further cooling, its integrated intensity increases, reaching a maximum around T*=70K, followed by a dramatic intensity reduction when the temperature is further decreased down to around 20K, as shown in **Fig. 1B** which confirms the previous X-ray scattering results [17].

Line cuts through the SDW-stripes peak along the h- and l-direction, are presented in the left and right panel of **Fig. 1C**, respectively. The solid lines correspond to results of Lorentzian profiles fitted to the data. The width of Lorentzian profiles gives the correlation lengths with the average value of $20$ nm (36 unit cells) in the NiO$_2$ plane while the average out-of-plane correlation is of the order is much smaller 2 nm (about 2 unit cells), reflecting the quasi two-dimensionality of the magnetic interactions [7,35].

In this work we have studied the temperature region of the decaying stripes signal intensity by recording micro X-ray diffraction maps at different temperatures. Three of such maps were collected by scanning the same sample area at 30K, 50K and T*=70K. At T* we observe the maximum SDW-stripes intensity. Our results are summarized in **Fig. 2**. In the spatial maps, red (blue) areas correspond to a high (low) SDW-stripes peak intensity. The inhomogeneity of the maps reveals regions with different stripe orientation or different degrees of order. We assign intermediate intensities (green-yellow) to the regions where sub-micron-size domains exist, which cannot be resolved by the 1 µm beam used in the experiment. **Fig. 2B** shows a statistical analysis of the SDW-stripes spatial distribution in terms of the probability density function $PDF(x = I/I_0)$ where $I_0$ is the average intensity of the map. At all temperatures, the distributions follow an exponentially truncated power-law behavior $PDF(x) = x^{-\alpha}exp(x/x_T)$ with a critical exponent $\alpha = 2.1 \pm 0.2$ and cut-off $x_T = 7 \pm 0.5$. Similar behavior has been reported for the distribution of the oxygen interstitials and the CDW order accompanied by local lattice distortions in the active layers of cuprates and related materials [2-4,30-34]. This underlines a spatial organization defining similar fractal-like geometries.

For investigating possible spatial rearrangements within the SDW-stripes pattern at the temperature of maximum ordering T*=70K we have calculated the radial distribution function $G(r)$ of the spatial maps (**Fig. 2C**). All the $G(r)$ curves show a similar exponential decay falling on the noise level at distances with $D_{SDW} \cong 10\ \mu m$. We associate this length to the size of the average SDW-stripes micrometric domain formed by ordered SDW-stripes nano-puddles, with the average 20 nm size. In fact the inhomogeneous aggregation of SDW-stripes puddles forms a complex fractal landscape of micron sized SDW-stripes rich domains illustrated in **Fig. 2D** separated by SDW-stripes poor domains as shown in panel A of Fig. 2.

The spatial arrangement of SDW-stripes puddles as a function of temperature shows a maximum density at T=70K with the formation of micron-scale striped pattern. At the low temperature (30K and 50K) the SDW-stripes puddles density decreases with large regions of disordered or rotated spin ordering. The here-determined temperature dependence of the SDW-stripes intensity is similar to the one reported for nickelates of different doping levels [17]. Thus, we can assume the observed behavior to be a general characteristic for SDW-stripes order. This behavior resembles what has been predicted for incommensurate cuprate stripes to occur at low temperatures, when a freezing to the lattice potential disturbs the long-range order [36-38]. We find intrinsic inhomogeneous spatial organization of SDW-stripes forming domains organized in micrometric puddles and a different spatial arrangement of SDW-stripes puddles as a function of temperature. The temperature of maximum SDW-peak intensity coincides with a reorganization of the SDW-stripes puddles of comparable size into a spatial pattern with regular inter-puddle distance. In conclusion, we have shown that the complexity of the investigated quantum striped material is different from but has similarities with superconducting quantum materials. The latter shows the coexistence of spin ordered, charge ordered and lattice ordered puddles in cuprates and related matter [39-41] pointing towards the possibility of quantum complex fluids in filamentary hyperbolic spaces [42] as it has been found in functional biological matter [43,44]. It is possible that in nickelate perovskites the complexity is driven by self-organization of polaronic localized spin or charge character [45,46] with associated tilts [47] forming quantum networks [49] driven by phase separation with anisotropic attractive interactions [50] controlled by anisotropic strain [51,52] and orbital degrees of freedom [26-29,53-55]. Our studies exploited novel high sensitive synchrotron radiation focused X-ray diffraction to shed new light on the formation of quantum spatial complexity [4-6,22-25,56,57] controlling new functionalities in quantum complex materials.

## ACKNOWLEDGMENTS

The Project was supported by the Helmholtz Virtual Institute: Dynamic Pathways in Multidimensional Landscapes. A.R. and G.C. acknowledge the Stephenson Distinguished Visitor Program by DESY. N.P. acknowledges the Italian Ministry for Education and Research and Marie Curie IEF project for partial financial support. A.A.N. acknowledges funding from Ministry of Research, Technology and Higher Education through Hibah WCU-ITB. Work in Cologne was supported by the DFG through SFB608 and by the German Ministry for Science and Education through contract 05K10PK2. Moreover, support by the DFG within SFB 925 and by the Superstripes-onlus is gratefully acknowledged. Work at the ALS, LBNL was supported by the Director, Office of Science, Office of Basic Energy Sciences, of the US Department of Energy (Contract No. DE-AC02-05CH11231).

**Author Contributions**:

A.R. conceived the project and designed all the experiments. C.S.L., M.S. and G.C. contributed to the planning of the experiments. The samples were grown by A.A.N., and characterized by M.Bu., C.T., C.S.L. and M.Br.; preliminary measurements using soft x-ray have been performed at ALS (Berkeley) by A.R., S.M., J.L., L.M. and S.R. and at P04-PETRA III (DESY) by M.Bu., C.S.L. and J.V..; micro X-ray diffraction experiments have been carried out at P10-PETRA III (DESY) by A.R., G.C., N.P., B.J., A.Z. and M.S.; data analysis has been done by A.R., G.C.. G.C., A.B., N.P., M.Br., C.S.L., A.R. and M.S. discussed the results and worked on the interpretation of the data. The manuscript has been written by A.R., G.C and A.B. collecting feedback from all the authors.

**Declaration of interests.** The authors declare no competing interest.

**REFERENCES**

[1] Dagotto, E. Complexity in strongly correlated electronic systems. Science **2005**, 309, 257-262

[2] Vojta, M. Lattice symmetry breaking in cuprate superconductors: stripes, nematics, and superconductivity. Advances in Physics **2009**, 58, 699-820.

[3] Bianconi, A., Saini, N. L. (Eds.). Stripes and related phenomena. Springer Science Business Media, **2001**.

[4] Campi, G. et al. Inhomogeneity of charge-density-wave order and quenched disorder in a high-Tc superconductor. Nature **2015,** 525, 359-362.

[5] Poccia, N. et al. Optimum inhomogeneity of local lattice distortions in La$_2$CuO$_{4+y}$. Proc. Natl. Acad. Sci. U.S.A. **2012,** 109, 15685-15690.

[6] Fratini, M. et al. Scale-free structural organization of oxygen interstitials in La$_2$CuO$_{4+y}$. Nature **2010,** 466, 841-844.

[7] Ulbrich, H. and Braden, M. Neutron scattering studies on stripe phases in non-cuprate materials. Physica C: Superconductivity **2012,** 481, 31-45 and references therein.

[8] Lee S.-H, Cheong, ., S-W. Phys. Rev. Lett. **1997,** 79, 2514

[9] Yoshizawa, H., Kakeshita, T., Kajimoto, R., Tanabe, T., Katsufuji, T., Tokura, Y. Stripe order at low temperatures in La2−xSrxNiO4 with 0.289≲x≲ 0.5. Physical Review B **2000**, 61(2), R 854 ().

[10] Kajimoto, R., Kakeshita, T., Yoshizawa, H., Tanabe, T., Katsufuji, T., Tokura, Y. Hole concentration dependence of the ordering process of the stripe order in La$_{2-x}$ Sr$_x$NiO$_4$. Physical Review B **2001**, 64(14), 144432.

[11] Lee, S.-H., Cheong, S.-W., Yamada, K., Majkrzak, C. F. Charge, canted spin order in La$_{2-x}$Sr$_x$NiO$_4$ (x=0.275,1/3). Phys. Rev. B **2001,** 63, 060405.

[12] Boothroyd, A. T., Freeman, P. G., Prabhakaran, D., Enderle, M., Kulda, J. Magnetic Order and Dynamics in Stripe-Ordered La2-xSrxNiO4 Physica B: Condensed Matter, **2004,** 345(1-4), 1-5.

[13] Freeman, P. G., Christensen, N. B., Prabhakaran, D., Boothroyd, A. T. The temperature evolution of the out-of-plane correlation lengths of charge-stripe ordered La1.725Sr0.275NiO4. In Journal of Physics: Conference Series **2010,** 200(1) 012037.

[14] Chen, C. H., Cheong, S-W., Cooper, A. S. Charge Modulations in La$_{2-x}$Sr$_x$NiO$_{4+y}$ Ordering of Polarons. Phys. Rev. Lett. **1993**, 71, 2461.

[15] Chow, K. H. et al. Muon-spin-relaxation studies of magnetic order in heavily doped La$_{2-x}$Sr$_x$NiO$_{4+\delta}$. Phys. Rev. B **1996,** 53, R14725.

[16] Schüßler-Langeheine, C., Schlappa, J., Tanaka, A., et al. . Spectroscopy of stripe order in La1.8Sr0.2NiO4 using resonant soft X-ray diffraction. *Physical Review Letters*, **2005,** *95*(15), 156402.

[17] Schlappa, J. et al. Static and fluctuating stripe order observed by resonant soft x-ray diffraction in La$_{1.8}$Sr$_{0.2}$NiO$_4$. preprint **2009** arXiv:0903.0994v1 .

[18] Coslovich, G., Huber, B., Lee, W. S., Chuang, Y. D., Zhu, Y., Sasagawa, T., ... Schoenlein, R. W., Ultrafast charge localization in a stripe-phase nickelate. Nature Cmmunications, **2013** 4, 2643.

[19] Wilkins, S. B., Hatton, P. D., Liss, K. D., Ohler, M., Katsufuji, T., & Cheong, S. W. High-resolution high energy X-ray diffraction studies of charge ordering in CMR manganites and nickelates. *International Journal of Modern Physics B*, **2000**, 14(29n31), 3753-3758.

[20] Ghazi, M. E. et al. Incommensurate charge stripe ordering in in La$_{2-x}$Sr$_x$NiO$_4$ for x= (0.33, 0.30, 0.275). Phys. Rev. B **2004,** 70, 144507.

[21] Du, C-H. et al. Critical Fluctuations and quenched Disordered Two-Dimensional Charge Stripes in La$_{5/3}$Sr$_{1/3}$NiO$_4$. Phys. Rev. Lett. **2000**, 84, 3911.

[22] Johnson, R. D., Mazzoli, C., Bland, S. R., Du, C. H., & Hatton, P. D.. Magnetically induced electric polarization reversal in multiferroic TbMn2O5: Terbium spin reorientation studied by resonant x-ray diffraction. *Physical Review B*, **2011**, *83*(5), 054438.

[23] Vecchini, C., Bombardi, A., Chapon, L. C., Lee, N., Radaelli, P. G., Cheong, S. W. Magnetic phase diagram and ordered ground state of GdMn$_2$O$_5$ multiferroic studied by x-ray magnetic scattering. Journal of Physics: Conference Series **2014,** 519, 012004.

[24] Pincini, D., Boseggia, S., Perry, R., Gutmann, M. J., Riccò, S., Veiga, L. S. I., ... Porter, D. G. Persistence of antiferromagnetic order upon La substitution in the Mott insulator Ca$_2$RuO$_4$. Phys. Rev. B, **2018,** 98(1), 014429.

[25] Price, N. W., Vibhakar, A. M., Johnson, R. D., Schad, J., Saenrang, W., Bombardi, A., ... & Radaelli, P. G. Strain Engineering a Multiferroic Monodomain in Thin-Film BiFeO3. *Physical Review Applied*, **2019,** *11*(2), 024035.

[26] Caprara, S., Sulpizi, M., Bianconi, A., Di Castro, C., Grilli, M. Single-particle properties of a model for coexisting charge and spin quasicritical fluctuations coupled to electrons. Phys. Rev. B, **1999,** 59(23), 14980

[27] Carlson, E. W., Yao, D. X., Campbell, D. K. (2004). Spin waves in striped phases. Physical Review B, **2004,** 70(6), 064505.

[28] Raczkowski, M. et al. Microscopic origin of diagonal stripe phases in doped nickelates. Phys. Rev. B **2006,** 73, 094429.

[29] Yamamoto, S., Fujiwara, T., Hatsugai, Y. Electronic structure of charge and spin stripe order in La$_{2-x}$Sr$_x$NiO$_4$ (x=1/3,1/2). Physical review B, **2007,** 76(16), 165114.

[30] Poccia, N. et al. Spatial inhomogeneity and planar symmetry breaking of the lattice incommensurate supermodulation in the high-temperature superconductor Bi$_2$Sr$_2$CaCu$_2$O$_{8+y}$. Phys. Rev. B **2011,** 84, 100504.

[31] Ricci, A. et al. Multiscale distribution of oxygen puddles in 1/8 doped YBa$_2$Cu$_3$O$_{6.67}$. Scientific Reports **2013,** 3, 2383.

[32] Ricci, A. et al. Networks of superconducting nano-puddles in 1/8 doped YBa$_2$Cu$_3$O$_{6.5+y}$ controlled by thermal manipulation. New J. Phys. **2014,** 16, 053030.

[33] Campi, G. et al. Scanning micro-x-ray diffraction unveils the distribution of oxygen chain nanoscale puddles in YBa$_2$Cu$_3$O$_{6.33}$. Phys. Rev. B **2013,** 87, 014517.

[34] Ricci, A. et al. Direct observation of nanoscale interface phase in the superconducting chalcogenide K$_x$Fe$_{2-y}$Se$_2$ with intrinsic phase separation. Phys. Rev. B **91**, 020503 (2015).

[35] Drees, Y. et al. Hour-glass magnetic excitations induced by nanoscopic phase separation in cobalt oxides. Nature Commun. **2014, 5**, 573.

[36] Mu, Y. and Ma, Y. Self-organizing stripe patterns in two-dimensional frustrated systems with competing interactions. Phys. Rev. B **2003, 67**, 014110 .

[37] Schmalian, J., Wolynes, P.G. Stripe glasses: Self-generated randomness in a uniformly frustrated system. Phys. Rev. Lett. **2000**, 85, 836

[38] Bogner, S., Scheidl, S. Pinning of stripes in cuprate superconductors. Phys. Rev. B **2001**, 64, 054517

[39] Poccia, N., Chorro, M., Ricci, A., Xu, W., Marcelli, A., Campi, G., Bianconi, A. Percolative superconductivity in La$_2$CuO$_{4.06}$ by lattice granularity patterns with scanning micro x-ray absorption near edge structure. Applied Physics Letters, **2014**, 104(22), 221903.

[40] Bianconi, A., Di Castro, D., Bianconi, G., Pifferi, A., et al. Coexistence of stripes and superconductivity: Tc amplification in a superlattice of superconducting stripes. Physica C: Superconductivity, **2000,** 341, 1719-1722.

[41] Chen, X., Schmehr, J. L., Islam, Z., Porter, Z., Zoghlin, E., Finkelstein, K., ... Wilson, S. D. Unidirectional spin density wave state in metallic (Sr$_{1-x}$La$_x$)$_2$IrO$_4$. Nature communications, **2018,** 9(1), 103.

[42] Campi, G., Bianconi, A. High-Temperature superconductivity in a hyperbolic geometry of complex matter from nanoscale to mesoscopic scale. Journal of Superconductivity and Novel Magnetism, **2016,** 29(3), 627-631.

[43] Campi, G., Di Gioacchino, M., Poccia, N., Ricci, A., Burghammer, M., Ciasca, G., Bianconi, A. Nanoscale correlated disorder in out-of-equilibrium myelin ultrastructure. ACS nano, **2017,** 12(1), 729-739.

[44] Campi, G., Cristofaro, F., Pani, G., et al. Heterogeneous and self-organizing mineralization of bone matrix promoted by hydroxyapatite nanoparticles. Nanoscale, **2017,** 9(44), 17274-17283.

[45] Zaanen, J., Littlewood, P. B. Freezing electronic correlations by polaronic instabilities in doped La$_2$NiO$_4$. Physical Review B, **1994,** 50(10), 7222.

[46] Kusmartsev, F. V., Di Castro, D., Bianconi, G., Bianconi, A., Transformation of strings into an inhomogeneous phase of stripes and itinerant carriers. Physics Letters A, **2000,** 275(1-2), 118-123.

[47] Gavrichkov, V. A., Shanko, Y., Zamkova, N. G., & Bianconi, A. Is There Any Hidden Symmetry in Stripe Structure of Perovskite High-Temperature Superconductors?. *J. Phys. Chem. Lett.*, **2019**, 10 (8), pp 1840–1844

[48] Bianconi, G. Quantum statistics in complex networks. *Physical Review E*, **2002**, *66*(5), 056123.

[49] Innocenti, D., Ricci, A., Poccia, N., Campi, G., Fratini, M., Bianconi, A. A model for liquid-striped liquid phase separation in liquids of anisotropic polarons. Journal of Superconductivity and Novel Magnetism, **2009,** 22(6), 529-533.

[50] Campi, G., Innocenti, D., Bianconi, A., CDW and similarity of the Mott insulator-to-metal transition in cuprates with the gas-to-liquid-liquid transition in supercooled water. Journal of Superconductivity and Novel Magnetism, **2015,** 28(4), 1355-1363.

[51] Agrestini, S., Saini, N. L., Bianconi, G., Bianconi, A. The strain of $CuO_2$ lattice: the second variable for the phase diagram of cuprate perovskites. Journal of Physics A: Mathematical and General, **2003,** 36(35), 9133.

[52] Bianconi, A., Agrestini, S., Bianconi, G., Di Castro, D., Saini, N. L. A quantum phase transition driven by the electron lattice interaction gives high Tc superconductivity. Journal of Alloys and Compounds, **2001,** 317, 537-541.

[53] Lichtenstein, A. I., Fleck, M., Oles, A. M., Hedin, L. Dynamical Mean-Field Theory of Stripe Ordering. In Stripes and Related Phenomena edited by Bianconi A. and Saini N.L., Springer, Boston, MA. **2002,** pp. 101-109.

[54] Raczkowski, M., Oleś, A. M. Competition between Vertical and Diagonal Stripes in the Hartree-Fock Approximation. In AIP Conference Proceedings **2003,** 678(1), 293-302.

[55] Oles, A. M. Charge and orbital order in transition metal oxides. Acta Phys. Polon. A **2010,** 118, 212.

[56] Beale, T. A. W., Wilkins, S. B., Johnson, R. D., Prabhakaran, D., Boothroyd, A. T., Steadman, P., ... & Hatton, P. D. Advances in the understanding of multiferroics through soft X-ray diffraction. *The European Physical Journal Special Topics*, **2012,** *208*(1), 99-106.

[57] Radaelli, P. G., & Dhesi, S. S. The contribution of Diamond Light Source to the study of strongly correlated electron systems and complex magnetic structures. *Philosophical Transactions of the Royal Society A: Mathematical, Physical and Engineering Sciences*, **2015,** *373*(2036), 20130148.

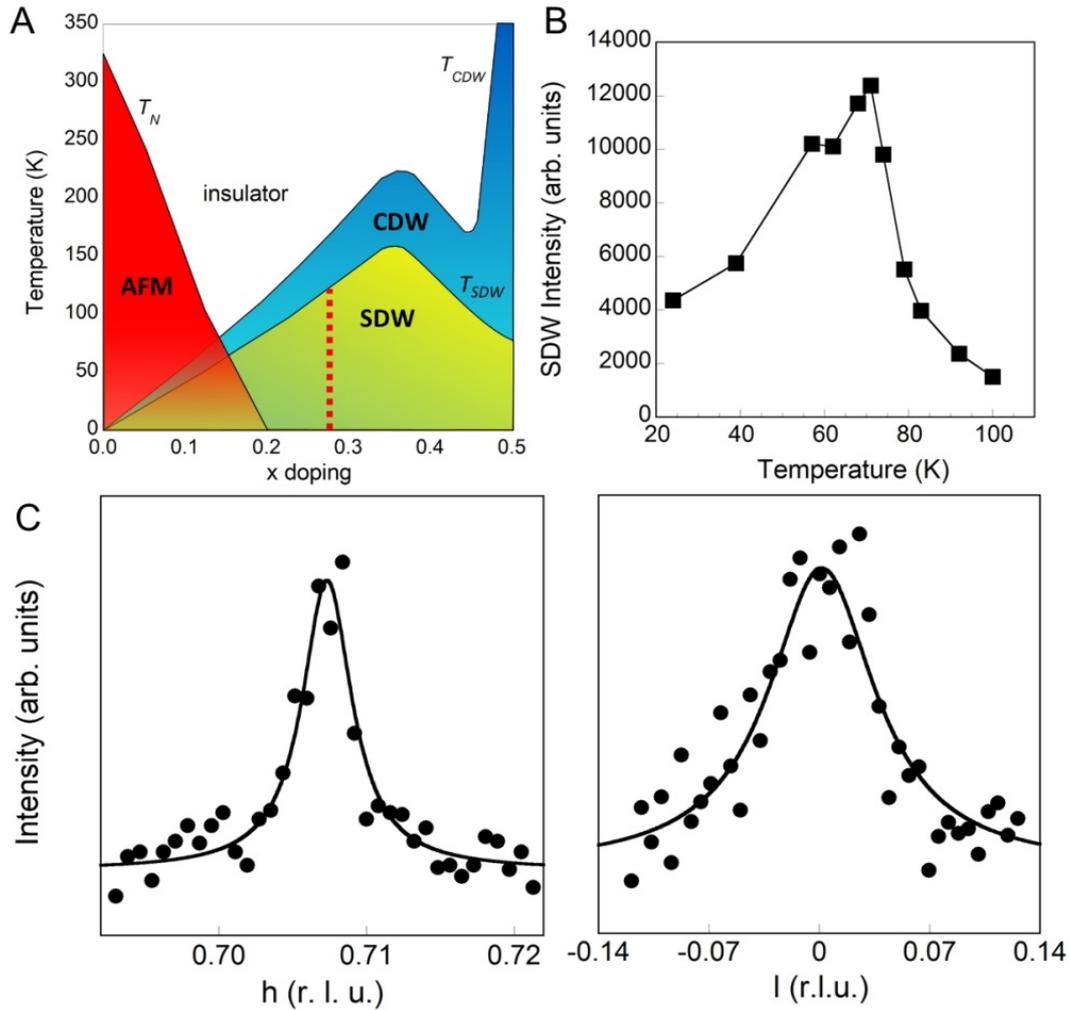

**Figure 1. Phase diagram of nickelate systems and Spin-Density-Wave order.**
**(A)** Temperature-doping Phase diagram of the nickelate systems. In red the insulating Antiferromagnetic order (AFM), in blue the charge-density-wave order (CDW-stripes) and in green the spin-density-wave order (SDW-stripes). The red dotted-line represents the temperature range where the sample of this work has been studied. **(B)** The intensity evolution of the SDW-stripes peak as a function of temperature. **(C)** SDW-stripes peak profile along the a* (left panel) and c* (right-panel) crystallographic directions, recorded at 30 K. The solid lines correspond to Lorentzian profiles fitted to the data, giving the in-plane and the out-of-plane correlation lengths around the average values of about 20 nm and 2 nm, respectively.

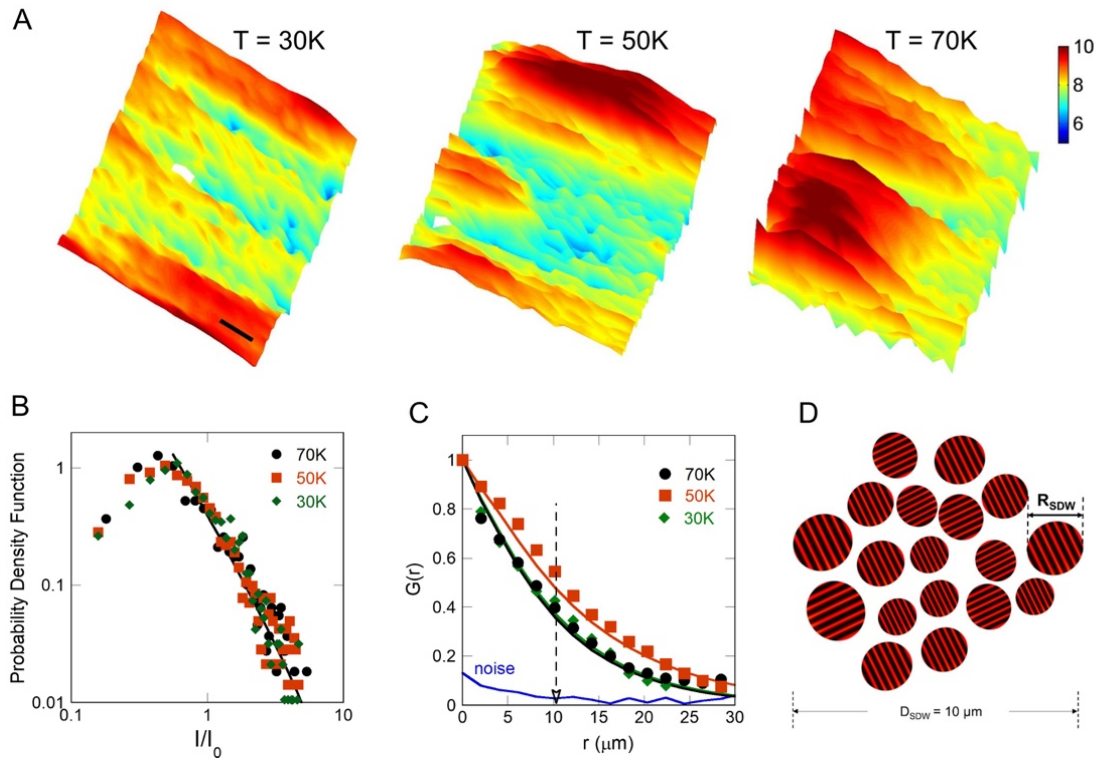

**Figure 2. Spatial inhomogeneous stripes order.**
**(A)** Each single pixel of the presented maps has been obtained recording the intensity of the XRD reflections probing SDW-stripes in a specific x,y position of the sample. In order to reconstruct the spatial maps, the sample has been scanned over an area of about $40 \times 80 \mu m^2$ in steps of $1 \mu m$ in both directions. The scale bar shown in the upper frame corresponds to $10 \mu m$. Red areas show SDW-stripes domains of the probed SDW-stripe orientation forming puddles of the order of about ten micrometers. Blue areas are representative of SDW-stripes domains with disordered or rotated SDW-stripes regions. **(B)** Probability density functions calculated from the stripes spatial maps intensities at 30K, 50K and T*=70K. The distributions rescale on the same curve and show evident fat-tails characterized by a power-law behavior with a critical exponent of 2.1 (solid black line). **(C)** The radial correlation function $G(r)$ calculated from the three spatial maps. The blue line represents the spatial correlation obtained for a random distribution of stripes peak intensities obtained by just shuffling the data at 70K. The $G(r)$ intensities decay exponentially, on the noise level, at D$_{SDW}$ indicating the size of a typical domain of 10 microns made by aggregation of individual nanoscale stripes puddles in the NiO$_2$ in the (a,b) plane. **(D)** Pictorial view of the stripes puddles of about R$_{SDW}$≈20 nm, given by the correlation length extracted from the width of the diffraction reflection lines (see Fig, 1), forming aggregates of about D$_{SDW}$≈10 micron size below the critical temperature T*. These aggregates show phase segregation with aggregates separated by regions with disordered or rotated stripes.